\documentclass{llncs}
\usepackage{amsfonts}
\usepackage[utf8]{inputenc}
\usepackage{graphicx}
\usepackage{xspace}
\usepackage{xcolor}
\usepackage{soul}
\usepackage{changepage}

\usepackage[]{algorithm2e}
\usepackage{amssymb}

\title{On the $k$-Hamming and $k$-Edit Distances}
\author{Chiara Epifanio\inst{1}\thanks{This research is funded by FONDO FINALIZZATO ALLA RICERCA DI ATENEO (FFR), University of Palermo, year 2023.} \and Luca Forlizzi \inst{2} \and Francesca Marzi \inst{2} \and Filippo Mignosi \inst{2}
\and Giuseppe Placidi \inst{3}\and Matteo Spezialetti \inst{2}} 

\institute{
DMI Department, University of Palermo, Italy, \email{chiara.epifanio@unipa.it},
\and
DISIM Department, University of L'Aquila, L'Aquila, Italy,\\
\email{\{luca.forlizzi, francesca.marzi, filippo.mignosi, matteo.spezialetti
\}@univaq.it}, \and MESVA Department, University of L'Aquila, L'Aquila, Italy,\\
\email{giuseppe.placidi@univaq.it}}

\begin{document}

\maketitle

\begin{abstract}
In this paper we consider the weighted $k$-Hamming and $k$-Edit distances, that are natural generalizations of the classical Hamming and Edit distances. As main results of this paper we prove that for any $k\geq 2$ the DECIS-$k$-Hamming problem is $\mathbb{P}$-SPACE-complete and  the DECIS-$k$-Edit problem is NEXPTIME-complete.
\end{abstract}

\section{Introduction}
\label{sec:introduction}

Measuring how dissimilar two strings are from each other, is a task that occurs often and which has great importance in various practical fields, such as biometric recognition and the study of DNA, up to spell checking.
A formal treatment of the problem passes through the definition of a notion of distance between strings.
Numerous distance functions have been proposed and studied from a computational point of view in the literature, based on the idea of measuring the minimum number of modification operations, chosen in a given set of admissible operations, necessary to transform one string into another: two of the best known are certainly the Edit distance and the Hamming distance, but since 1950 other distances have been introduced and scientific studies have been carried on (cf. for instance 
\cite{hamming1950error,damerau1964technique,levenshtein1966binary,wagner1974string,wagner1975complexity,wagner1975extension,Bergroth00,amir2006swap,CormodeM07,Meister15,BackursI18,BarbayP18,ACFGMMDlt23,AFMWords21,AFMWords21-TCS,beal2022checking,faro2018efficient,anselmo2023hypercubes}).
In this framework, measuring how similar two strings are is then formalized as an optimization problem, i.e. minimizing the amount of operations to transform one into the other. It is quite useful to also consider the decision version of one of such problem, in the following way: in any instance there are two words together with a natural number $h$ and we ask whether or not the two given words have a distance (Hamming, edit or another one) that is smaller than or equal to $h$. 

Previous approach could seem well formalized but there is something hidden: is the description of the fixed distance (Hamming, edit or another one) \emph{included} inside the instances of the problem \emph{or} the description of the distance has to be considered as a constant that can vary depending on the problem but that should not be considered in the asymptotic analysis of the algorithms that solves the problems?

Usually the second approach is the one that seems preferred in literature. For instance we say that the complexity of the classical algorithm for the edit distance is $O(nm)$, where $n$ and $m$ are the lengths of the two strings. And, moreover, it seems that there is no much difference if we choose the first approach for the edit distance. 

On the contrary something different happens in some cases. In \cite{wagner1974string} it is proved that including the description of a special distance inside the instances gives rise to an $\mathbb{NP}$-hard problem, whilst it has been proved much later in \cite{Meister15,BarbayP18} that the same problem has a solving algorithm that is polynomial when the size of the description of the distance is considered as a constant. The interested reader can see \cite{BarbayP18} and references therein for more details.

In this paper we study the problems of computing the $k$-Hamming and the $k$-Edit distances, for $k\geq 2$, in the first setting, i.e. we suppose that the description of the distance is a part of the instances. The study of these problems following the second approach is still open, as discussed in Section \ref{sec:conclusions}.

In Section~\ref{preliminaries} we introduce our notation and some formal definitions. 
Section~\ref{complexity} is devoted to prove that, for $k\geq 2$, the decision problems of computing the $k$-Hamming (DECIS-$k$-Hamming) and the $k$-Edit (DECIS-$k$-Edit) are, respectively, $\mathbb{P}$-SPACE-complete and NEXPTIME-complete. To do so, we follow the same strategy for both problems. 
First we prove the results for $k=3$, using polynomial time reductions from any $L\in\mathbb{P}$-SPACE to DECIS-$3$-Hamming and from any $L\in$ NEXPTIME to DECIS-$3$-Edit, and straightforwardly extend them to larger values of $k$. Then we reduce (in polynomial time) the problems with $k=3$ to the respective problems with $k=2$, proving the results also for these cases.
Section \ref{sec:conclusions} concludes the paper foreshadowing possible research developments.

\section{Preliminaries}\label{preliminaries}

Given a finite alphabet $\Sigma$ of cardinality $\sigma$, a string over $\Sigma$ is a sequence $w= w_1w_2\ldots w_n$, with $w_i\in\Sigma$, for any $1\leq i\leq n$. The number of characters composing a string $w$ is called its \emph{length}, denoted by $|w|=n$. The string of length 0, also called the empty string, is indicated by $\epsilon$.
We denote by $\Sigma^*$ the set of all strings on $\Sigma$ and by $\Sigma^n$ the set of all strings of length $n$ in $\Sigma^*$. Trivially $\epsilon \in \Sigma^*$, for any $\Sigma$.
String $x$ is a substring of string $w$ if there exist $u$ and $v$ such that it is possible to write $w$ as the concatenation of $u$, $x$ and $v$ i.e., such that $w=uxv$. The empty string is a substring of any string.
We denote by $x=w_j \ldots w_{j+k-1}$ the substring of $w$ of length $k$ that appears in $w$ at position $j$, i.e. $w=uxv$, with $|u|=j-1$, $|x|=k$ and $|v|=|w|-(k+j-1)$. 

Given a string $v$, it is possible to define a set $Op\{o:\Sigma^*\rightarrow \Sigma^*\}$ of operations that allow to modify it in a new string $w$. Some well-studied subsets of operations are the {\em edit operations}:
\begin{definition}Given a string $v\in\Sigma^*$, we define:
    
\begin {itemize}
\item \textsc{Insertion (I, $\epsilon \rightarrow a$)} allows to insert a character $ a \in \Sigma $ in any position $i$ of $ v $, i.e. $w=v_1 \ldots v_{i-1} a v_i \ldots v_{|v|}$;
\item \textsc{Deletion (D, $a \rightarrow \epsilon$)} is the removal of any character $ a=v_i $ in $ v $, i.e. $w=v_1 \ldots v_{i-1} v_{i+1} \ldots v_{|v|}$;
\item \textsc{Substitution (S, $a\rightarrow b$)} replaces any character $ a=v_i$ in $ v $ with another character $ b \in \Sigma $ in the same position, i.e. $w=v_1 \ldots v_{i-1} b v_{i+1} \ldots v_{|v|}$.
\end{itemize} 
\end{definition}

We describe here some other operations that allow to define some more distances.

\begin{definition}
Given a string $v\in \Sigma^*$, we define the following operations on $v$.

\begin{itemize}
\item \textsc{2-Substitution (2S, $a_1a_2\rightarrow b_1b_2$)} replaces any pair of consecutive characters $ a_1a_2= v_iv_{i+1}$ in $v$, with another pair of characters $ b_1b_2 \in \Sigma^2 $, i.e. $w=v_1 \ldots v_{i-1} b_1b_2 v_{i+2} \ldots v_{|v|}$.
 \item \textsc{$k$-Substitution ($k$S, $a_1\ldots a_k\rightarrow b_1 \ldots b_k$)} allows substitutions of $ k $ consecutive characters all at once. It replaces in $v$ the substring $a_1\ldots a_k=v_i \ldots v_{i+k-1}$ with $b_1 \ldots b_k$, i.e. 
$w=v_1 \ldots v_{i-1} b_1 \ldots b_k v_{i+k} \ldots v_{|v|} $. Obviously, $kS$ is a generalization of the $S$ and $2S$ previously introduced, e.g. $S=kS$ if $k=1$. 
\end{itemize} 
 
\end{definition}

For the sake of readability, we henceforth use the notation $Op=\{A_1,\ldots ,A_m\}$, with $A_1,\ldots ,A_m \in \{I,D,kS |\ k\in \mathbb{Z}^+\}$ to indicate that all the possible operations defined by each $A_i$ are in $Op$, e.g. $Op=\{I\}$ allows all the insertions $\epsilon \rightarrow a$ with $a\in\Sigma$.

At this point, we can define a {\em cost} function $\gamma: \Sigma^* \times \Sigma^* \rightarrow \mathbb{Z}^+ $ for each operation. That cost can be fixed or can depend on the type of operation or on the characters on which it is applied.

\begin{definition}\label {def: distance}
Let $v,w \in \Sigma^*$ be two strings, $Op$ be a set of operations defined on $\Sigma^*$, $\gamma $ be an arbitrary cost function. 
If $T=t_1t_2\ldots t_p$ is a sequence of operations over $Op$, i.e. $T(v)=t_p(\ldots(t_2(t_1(v)))\ldots)$, the overall cost of the sequence is:
$$\gamma(T)=\sum_{i=1}^p \gamma(t_i).$$  The \textit{distance} between $v$ and $w$ is the minimum cost  required to transform $v$ into $w$ through operations in $Op$, i.e.
\begin{equation}
\delta (v, w) = \min \{ \gamma(T) \mid T(v) = w\}.
\end{equation}
\end{definition}

Depending on the set $Op$ of operations allowed on $\Sigma^*$ we can define different distances.
\begin{definition}
The \textit{Edit distance} between $v$ and $w$ is $\delta(v,w)$, considering $Op = \{I, D, S\}$. \end{definition}

The Edit distance is also formally known as \emph{Levenshtein distance}, due to the work carried out by Vladimir Levenshtein who introduced for the first time an algorithmic approach to calculate this distance \cite {levenshtein1966binary}. 

\begin{definition}We define \textit{Hamming distance} between $v$ and $w$ \cite{hamming1950error} $\delta(v,w)$, when $Op=\{S\}$. 
\end{definition}

Apart from the well-studied Edit Distance and Hamming Distance, it is possible to define some other distances between strings.

\begin{definition} [$2$-Edit Distance]\label{def: 2-edit_dist}
The \emph{$2$-Edit distance} between two strings $ v $ and $ w $ is the minimum cost to transform the string $ v $ into $ w $, $\delta(v,w)$, setting the set of admissible operations $ Op = \{ I, D, S, 2S \} $. It is a direct extension of the previously defined Edit distance, with the addition of the double substitution operation.
\end{definition}

Last but not least we define the generalizations of $2$-Edit and Hamming distances, the $k$-Edit and $k$-Hamming distance, respectively, for a given $k\geq 2 \in \mathbb{Z}$.

\begin{definition} \label{def: K-Hamming_dist}
Given two strings $v$ and $w$ and a positive integer $k\geq 2$, 

\begin{itemize}
    \item \normalfont{\textbf{$k$-Edit distance}} the \emph{$k$-Edit distance} between $v$ and $w$ is $\delta(v,w)$, with $Op=\{I,D,S,k$S$\}$.  
    \item \normalfont{\textbf{$k$-Hamming distance}} the \emph{$k$-Hamming distance} between $v$ and $w$ is $\delta(v,w)$, with $Op=\{k$S$\}$. 

\end{itemize}
\end{definition}

\section{Complexity}\label{complexity}

\subsection{DECIS-$3$-Hamming $\mathbb{P}$-SPACE completeness}
\label{sub:3H}
We prove in this section that DECIS-$3$-Hamming problem is $ \mathbb {P} $ - SPACE-complete.
DECIS-$3$-Hamming contains all the strings encoding quadruples of the form $<v, w, D, h>$ where $v$ and $w$ are two strings on $\Sigma^n$ of the same length $n$, $D$ is an encoding string that describes the weighted $3$-Hamming distance we are considering and $h$ is an integer.

Hence, a istance $x =< (v, w, D, h) >$ fits into DECIS-$3$-Hamming if and only if $D(v,w)\leq h$. Therefore
$$\mbox{DECIS-}3\mbox{-Hamming} = \{< (v, w, D, h) > : D(v,w) \leq h\}.$$

In order to say that DECIS-$3$-Hamming is $\mathbb{P}$-Space complete, we need to prove the two following properties: a) DECIS-$3$-Hamming is in $\mathbb{P}$-Space; b) for every language $L$ in $\mathbb{P}$-Space there exists a polynomial reduction from $L$ to DECIS-$3$-Hamming.

\begin{theorem}
    \item DECIS-$3$-Hamming is in $\mathbb{P}$-Space.
\end{theorem}
\begin{proof}
By a corollary to Savitch's Theorem \cite{SAVITCH1970177} we know that $\mathbb{P}$-Space=$\mathbb{NP}$-Space.

Hence, proving that the problem is in $\mathbb{NP}$-Space will be enough to prove the Theorem.

We define a Nondeterministic Turing Machine $N$ that accepts the DECIS-$3$-Hamming language in polynomial space, even in the worst case. $N$ starts with $<v,w,D,h>$ coded on its tape and operates iteratively. In each loop, it non-deterministically chooses a substitution to apply to the string, executes it and updates $h$ by subtracting the weight of the substitution just chosen. $N$ exits the while loop when $v$ becomes equal to $w$ or $h$ is negative. In both cases it will be possible to establish whether the given instance belongs to DECIS-$3$-Hamming. It is possible to observe that the total occupied space is linear with respect to the length of the input strings, hence DECIS-$3$-Hamming is in $\mathbb{NP}$-SPACE, and, therefore, in $\mathbb{P}$-SPACE.

\begin{algorithm}[H]
		\caption{Algorithm followed by $N$ for solving DECIS-$3$-Hamming}
		\label{alg:ntm_3-ham}
\begin{center}\small
 \fbox{
  \begin{minipage}{22cm}
   \begin{tabbing}
 xxx \= xxx \= xxx \= xxx \= xxx \= xxx \=\kill
1. \> {\bf begin}\\
2. \> \> {\bf while} $v \neq w \wedge h \geq 0$\\
3. \> \> \> non-deterministically choose a substitution to apply;\\
4. \> \> \> apply the substitution to string $v$;\\
5. \> \> \> subtract the weight of the substitution from $h$;\\
6. \> \> {\bf if} $v==w \wedge h \geq 0$\\
7. \> \> \> {\bf ACCEPT}\\
8. \> \> {\bf else}\\
9. \> \> \> {\bf DO NOT ACCEPT}\\
10.\> {\bf end}
   \end{tabbing}
  \end{minipage}
 }
\end{center}
\end{algorithm}

\bigskip

\end{proof}

\begin{theorem}\label{th:rid_PS_D3H}
For each language $L$ in $\mathbb{P}$-SPACE there is a polynomial time reduction from $L$ to DECIS-$3$-Hamming.
\end{theorem}

\begin{proof} 
If $L$ is in $\mathbb{P}$-SPACE there exists a deterministic Turing machine $$M=<Q,\Gamma, B, \Sigma,\Delta,q_0,F>$$ that stops on every input of size $n$ in $O(c^{q(n)})$ time and decides $L$ in polynomial space $O(p(n))$, being $c$ a constant and $p$ and $q$ two polynomials.
 
The idea is to take a Semi-Thue system \cite{davis1994computability} that simulates $M$ and make it a $3$-Hamming distance by adding all the missing substitutions with a very large weight, leaving instead with a very low weight, i.e. equal to $1$, the substitutions of the Semi-Thue system.

We define $M'$ as the Turing Machine that accepts the DECIS-$3$-Hamming language.
We define an algorithm for mapping each instance $x$ in $L$ into an instance $x' = <(v,w,D,h)>$, such that $M$ accepts $x$ if and only if $M'$ accepts $x'$.
We formally define the parameters of instance $x'$ as follows.

\begin{itemize}
    \item $v = \$B^{p(n)+1} q_0 x B^{p(n)+1}\$$, where $\$ \notin \Gamma$;
\item $w = \$B^l\$$ with $l = 2p(n)+n+3$;
\item $h =\min \{c^m > c^{q(n)} + 2p(n)+4+n\}$.
This value of $h$ can be represented in base $c$ as the string obtained by the concatenation of $1$ and $m$ times $0$, with $m=\lceil \log_c c^{q(n)} + 2p(n)+4+n \rceil.$ 
\end{itemize}

The last parameter to define is the distance $D$. We note immediately that the description of the distance is independent of $x$, therefore it is constant with respect to $n$. This distance is a weighted 3-Hamming that assumes only two weights $1$ and $h+1$. To give the full description of $D$ we would need to define the weight for all $3$-substitutions. For each $y \in \Gamma$ the following $3$-substitutions with cost $1$ are produced:
\begin{itemize}
\item every transition $\Delta(q_h,a) = (q_j,b,R)$ in $M$ produces $yq_ha \rightarrow ybq_j$ in $D$;

\item every transition $\Delta(q_h,a) = (q_j,b,L)$ in $M$ produces $yq_ha \rightarrow q_jyb$ in $D$; 

\item every transition $\Delta(q_h,B) = (q_j,b,R)$ in $M$ produces $yq_hB \rightarrow ybq_j$ in $D$;

\item every transition $\Delta(q_h,B) = (q_j,b,L)$ in $M$ produces $yq_hB \rightarrow q_jyb$ in $D$.

\end{itemize}

In addition, the following $3$-substitutions with cost $1$ are added, for each $q_s\in F$ and $a,b \neq \$$, with $\#_l, \#_r \notin \Gamma$.

\begin{itemize}
\item $aq_sb \rightarrow \#_lB\#_r$
\item $a\#_lB \rightarrow \#_lBB$
\item $\$\#_lB \rightarrow \$BB$
\item $B\#_rb \rightarrow BB\#_r$
\item $B\#_r\$ \rightarrow BB\$$
\end{itemize}

This set of 3-substitutions is required if a $q_s \in F$ appears on the simulated tape. In fact, it is used to erase the entire tape.
For the remaining undefined 3-substitutions we set the cost to $h+1$.
\end{proof}

It is possible to observe that the algorithm is polynomial.

\begin{theorem}
Let $x$ be an instance in $L\in \mathbb{P}$-SPACE, the $x$ transformation in $x'$ just defined is a reduction, i.e. $$ x\in L \iff x' \in \mbox{ DECIS-}3\mbox{-Hamming}.$$
\end{theorem}

\begin{proof} Suppose first that $x\in L$. This means that there exists a finite sequence of ID $\alpha_1 \ldots \alpha_t$ such that $\alpha_1=q_1x$, for any $i<t<c^{q(n)}$ $\alpha_i\vdash\alpha_t$ and $\alpha_t$ is a final ID.
For each implication from one ID to another one there is a corresponding transition rule which can be simulated by a substitution of unit weight in the distance $D$, as previously described.
Formally we match $\alpha_i$ to the string $v$ and at the end of the simulation we will have reached $\alpha_t$ which will correspond to a string $v'$ containing $q_s$. In this way we will be able to say that there exists a sequence of substitutions of unitary weight in $D$ which, starting from $v$, allows us to arrive at $v'$ with a total weight less than $c^{q(n)}$. Using, at this point, the substitutions of unitary weight that cancel the symbols different from $\$$ and $B$ around $q_s$ we will obtain the string $w=\$B^l\$$, with $l=2p (n)+n+3$. In total, therefore, the cost of obtaining $w$ is less than or equal to $c^{q(n)}+2p(n)+n+4$ and therefore less than $h$. So $x' \in$ DECIS-$3$-Hamming.

\bigskip
Let us prove now the converse. We do it by contraposition.
If $x \not \in L$ then there is no sequence of transitions that can lead the initial ID to an ID in which a final state appears. In the simulation using $3$-substitutions, no sequence of substitutions of unitary weight can ever transform the string $v$ into a string $v'$ containing a final state and therefore $w$ cannot be obtained.
The only way to get an accepting state on the tape would be to use a substitution costing $h+1$. But in this case the $3$-Hamming distance between $v$ and $w$ will certainly be greater than $h$, so $x'\not \in $ DECIS-$3$-Hamming.

\end{proof}

\begin{theorem}\label{th:khamm}
Any DECIS-$k$-Hamming, with $k\geq 3$, is $\mathbb{P}$-SPACE-complete.
\end{theorem}

\begin{proof}
    It is easy to observe that the previous proof can be used to demonstrate, by induction, the $\mathbb{P}$-SPACE-completeness of any DECIS-$k$-Hamming problem, with $k\geq 3$, since: a) Algorithm \ref{alg:ntm_3-ham} works for any DECIS-$k$-Hamming; b) There exists a polynomial time reduction from DECIS-$k$-Hamming to DECIS-$(k+1)$-Hamming ($k\geq 2$). The reduction has just to pad input and target string (to handle strings with lenght $3$) and to inhibit any $(k+1)$-substitution that does not represent a $k$-substitution.
\end{proof}

\subsection{DECIS-$2$-Hamming $\mathbb{P}$-Space-Completeness}
\label{sub:2H}
In this section we prove that also DECIS-$2$-Hamming is $\mathbb{P}$-Space Complete. We first define the following set, for any $k\in \mathbb{Z}^+$ and $x,y \in \Sigma^k$:
$$\mbox{DECIS'-}k\mbox{-Hamming}= \{<v,w,D,h>|\delta(v,w)\leq h, \gamma(x\rightarrow y)\in \{1, h+1\}\}$$

It is to note that the proof of $\mathbb{P}$-Space-completeness of DECIS-$3$-Hamming holds for DECIS'-$3$-Hamming, since: a) DECIS'-$3$-Hamming is a special case of DECIS-$3$-Hamming, thus the Algorithm \ref{alg:ntm_3-ham} is valid; b) the reduction defined in Theorem \ref{th:rid_PS_D3H} actually produces instances of DECIS'-$3$-Hamming.

We can, therefore, state the following lemma.
\begin{lemma}
 DECIS'-$3$-Hamming is $\mathbb{P}$-Space-Complete.     
\end{lemma}

It is also to note that an algorithm similar to Algorithm \ref{alg:ntm_3-ham} can be defined for DECIS-$2$-Hamming, thus:
\begin{lemma}
    DECIS-$2$-Hamming $\in \mathbb{P}$-Space.
\end{lemma}

\begin{lemma}
  There is a reduction from DECIS'-$3$-Hamming to DECIS-$2$-Hamming.     
\end{lemma}
\begin{proof} It is possible to prove this reduction thanks to a technique which belongs to the folklore of Information Theory and to Markov chains. This technique reduces the dependence of a random variable on $k$ previous random variables, including itself, to just two random variables, including itself, via a sliding window over a larger alphabet.

Let $x=<v,w,D,h>$ be an instance in DECIS'-$3$-Hamming, we transform it in an instance $x'=<v',w',D',h'>$ in DECIS-$2$-Hamming, where
\begin{itemize}
    \item $v'=c_1c_2 \ldots c_{n+1}$ is obtained from $v=a_1a_2\ldots a_n$, by:
    \begin{itemize}
        \item $\$$-padding $v$, i.e. $\bar{v}=b_1b_2 \dots b_n+2=\$v\$$;
        \item coding any symbol of $v'$ as a pair of consecutive symbols of $\bar{v}$, obtained with a sliding window of length $2$ and stride $1$, i.e. $c_i=(b_i,b_{i+1})$
    \end{itemize}
    \item $w'$ is constructed from $w$ in an analogous way;
    \item $h'=3h$;
    \item for each $3$-substitution $abc \rightarrow def$, with $\gamma=1$ in $D$, the following unit cost $2$-substitutions are added to $D'$:
    \begin{itemize}
        \item $(ab)(bc)\rightarrow S_{(ab)(de)}^\leftarrow S_{(bc)(ef)}^\rightarrow$;
        \item $(xa)S_{(ab)(de)}^\leftarrow \rightarrow (xd)(de)$, $\forall x\in \Sigma \cup \{\$\}$
        \item $S_{(bc)(ef)}^\rightarrow (cx)\rightarrow (ef)(fx)$, $\forall x\in \Sigma \cup \{\$\}$
    \end{itemize}
    \item any other $2$-substitution has cost $h'+1$
\end{itemize}

The algorithm is a polynomial in the size of the input, indeed: a) $|v'|=|v|+1$ and $|w'|=|w|+1$; b) coding $h'=3h$ requires linear time; c) the algorithm increases the size of the alphabet with a polynomial function and coding $D'$ requires $O(|\Sigma'|^2)$ steps.

Moreover, it is possible to observe that the algorithm is a reduction, i.e.: $$ x\in \mbox{ DECIS'-}3\mbox{-Hamming} \iff x' \in \mbox{ DECIS-}2\mbox{-Hamming}$$

Suppose $x\in \mbox{ DECIS'-}3\mbox{-Hamming}$, i.e. $\exists T=t_1 \ldots t_k\ s.t.\ T(v)=w, \ \gamma(T)\leq h$, with each $t_i \in D$. Then, $\exists T'=t'_1 \ldots t'_{3k}\ s.t.\ T'(v')=w', \ \gamma(T')\leq h'$, with each $t'_i \in D'$. $T'$ is obtained by $T$, by translating each $t_i$ into the corresponding sequence of $2$-substitutions described by the algorithm, thus $$x\in \mbox{ DECIS'-}3\mbox{-Hamming} \Rightarrow x' \in \mbox{ DECIS-}2\mbox{-Hamming}$$

Suppose $x\notin \mbox{ DECIS'-}3\mbox{-Hamming}$, i.e. $\forall T=t_1 \ldots t_k\ s.t.\ T(v)=w, \ \gamma(T)> h$, with each $t_i \in D$. Since the algorithm, by construction, do not insert any $S_{(ab)(de)}^\leftarrow$ or $S_{(bc)(ef)}^\rightarrow$ symbols in $w'$, the only way to obtain $w'$ from $v'$ is to remove all these symbols from the string, thus completing simulated (and legal) $3$-substitutions in the input instance. Therefore, $\forall T'\ s.t.\ T'(v')=w', \ \gamma(T')>3h$, thus $$x\notin \mbox{ DECIS'-}3\mbox{-Hamming} \Rightarrow x' \notin \mbox{ DECIS-}2\mbox{-Hamming}$$

\end{proof}

These lemmas imply the following result.
\begin{theorem}
    Decis-$2$-Hamming is $\mathbb{P}$-Space-Complete.
\end{theorem}

\subsection{DECIS-$3$-Edit NEXPTIME-completeness}
\label{sub:3E}
\label{sec:d3e-nexptime}
We will now prove that DECIS-$3$-Edit distance is NEXPTIME-complete, that is: a) DECIS-$3$-Edit $\in$ NEXPTIME; b) $\forall L \in$ NEXPTIME, there exists a polynomial time reduction from $L$ to DECIS-$3$-Edit.

\begin{theorem}
DECIS-$3$-Edit $\in$ NEXPTIME
\end{theorem}
\begin{proof}
    We show a Nondeterministic Turing Machine $N$ that, given $x =< (v, w, D, h) >$ in input, accepts if and only if $D(v,w)<h$. $N$ acts as described in \ref{alg:ntm_3-edit}.

\begin{algorithm}[H]
		\caption{Algorithm followed by $N$ for solving DECIS-$3$-Edit}
		\label{alg:ntm_3-edit}
\begin{center}\small
 \fbox{
  \begin{minipage}{22cm}
   \begin{tabbing}
 xxx \= xxx \= xxx \= xxx \= xxx \= xxx \=\kill
1. \> {\bf begin}\\
2. \> \> {\bf while} $v \neq w \wedge h \geq 0$\\
3. \> \> \> non-deterministically choose an edit operation $o$ to apply;\\
4. \> \> \> apply $o$ to string $v$;\\
5. \> \> \> subtract $\gamma(o)$ from $h$;\\
6. \> \> {\bf if} $v==w \wedge h \geq 0$\\
7. \> \> \> {\bf ACCEPT}\\
8. \> \> {\bf else}\\
9. \> \> \> {\bf DO NOT ACCEPT}\\
10.\> {\bf end}
   \end{tabbing}
  \end{minipage}
 }
\end{center}
\end{algorithm}
Since $\gamma: \Sigma^* \times \Sigma^* \rightarrow \mathbb{Z}^+$, the algorithm performs at most $h=O(2^n)$ loops, each composed by linear time operations. Thus, $M$ halts in an exponential time in $n$ and DECIS-$3$-Edit $\in$ NEXPTIME.
\end{proof}

\begin{theorem}
$\forall L \in$ NEXPTIME, there exists a polynomial time reduction from $L$ to DECIS-$3$-Edit
\end{theorem}

\begin{proof}
    If $L\in$ NEXPTIME, there exists a Nondeterministic Turing Machine $$N'=<Q,\Gamma, B, \Sigma,\Delta,q_0,F>$$ that recognizes if $x\in L$ and stops within an exponential number of moves, i.e. if $n=|x|$, it will halt after $2^{p(n)}$ steps at most, where $p(n)$ is a polynomial function of $n$. The reduction transforms any instance $x$ for $L$ in an instance $x' =< (v, w, D, h) >$ for DECIS-$3$-Edit as follows:
    \begin{itemize}
        \item $v=\$q_0x\$$, with $\$\notin\Gamma$
        \item $w=\$\$$
        \item $h=5*2^{p(n)}+2*(n+1)$
    \end{itemize}
    Finally, $D$ is defined in the following way:
    \begin{enumerate}
        \item \label{it:1}any insertion has cost $\gamma=h+1$, with the exception of $\epsilon \rightarrow B_1$ (being $B_1 \notin \Gamma$ a new blank symbol), that has cost $\gamma=1$; 
        \item \label{it:2}any deletion has cost $\gamma=h+1$, with the exception of $*\rightarrow\epsilon$, that costs $\gamma=1$, where $*\notin \Gamma$ is a new symbol used to delete the simulated tape after the acceptance of $N'$;
        \item any substitution has cost $\gamma=h+1$
        \item any $3$-substitution has cost $\gamma=h+1$, with the following exceptions:
        \begin{enumerate}
            \item for each element of $\{<(q,a),(p,b,R)>|(p,b,R)\in \Delta(q,a)\}$, with $q$ and $p$ state symbols not in $\Gamma$ and $a,b \in \Gamma$:
            \begin{enumerate}
                \item \label{it:4.1.1} $qax\rightarrow bpx$, with $\forall x \in \Gamma$, has cost $\gamma=3$;
                \item \label{it:4.1.2}$qa\$\rightarrow bp\$$, with $\forall p \notin F$, has cost $\gamma=1$;
                \item \label{it:4.1.3}$qa\$\rightarrow bp\$$, with $\forall p \in F$, has cost $\gamma=3$;
            \end{enumerate}
            \item for each element of $\{<(q,a),(p,b,L)>|(p,b,L)\in \Delta(q,a)\}$, with $q$ and $p$ state symbols not in $\Gamma$ and $a,b \in \Gamma$:
            \begin{enumerate}
                \item \label{it:4.2.1}$xqa\rightarrow pxb$, with $\forall x \in \Gamma$, has cost $\gamma=3$;
                \item \label{it:4.2.2}$\$ qa\rightarrow p\$ b$, with $\forall p \in Q$, has cost $\gamma=1$;
            \end{enumerate}
            \item to simulate moves that require to expand the tape length behind $|x|$, the following $3$-substitutions have cost $\gamma=1$:
            \begin{enumerate}
                \item \label{it:4.3.1} $qB_1\$ \rightarrow qB\$$;
            \item \label{it:4.3.2} $p\$ B_1 \rightarrow \$ pB$;
            \end{enumerate}
            \item to delete symbols and reach the target string $\$\$$ after the acceptance $N'$, with $p \in F$ and $a,b \in \Gamma$, the following $3$-substitutions have cost $\gamma=1$:
            \begin{enumerate}
                \item \label{it:4.4.1}$apb \rightarrow \#_l * \#_r$;
                \item \label{it:4.4.2}$\$pa \rightarrow \$ * \#_r$;
                \item \label{it:4.4.3}$ap\$ \rightarrow \#_l*\$$;
                \item \label{it:4.4.4}$\$p\$ \rightarrow \$ * \$$;
                \item \label{it:4.4.5}$a\#_l* \rightarrow \#_l * *$;
                \item \label{it:4.4.6}$*\#_ra \rightarrow **\#_r$;
                \item \label{it:4.4.7}$\$\#_l* \rightarrow \$**$;
                \item \label{it:4.4.8}$*\#_r\$ \rightarrow **\$$.
            \end{enumerate}
        \end{enumerate}
    \end{enumerate}
The algorithm takes polynomial time $q(n)$: writing $v$ and $w$ requires linear time in $n$, while coding $D$ would take $O(|\Gamma|^6)$. Moreover, it is actually a reduction, i.e.: $$ x\in L \iff x' \in \mbox{ DECIS-}3\mbox{-Edit.}$$

    Suppose $x \in L$. There exists a finite sequence of non-deterministic moves (and, therefore, of IDs) that makes $N'$ accept $x$. It is easy to see that there is a corresponding sequence of transformations that modifies $v$ and results in the string $\$xpy\$$, with $x,y \in \Gamma^*$ and $p\in F$. Each $3$-substitution that simulates a $N'$ move has cost $\gamma=3$, if it does not involve the $\$$ symbol (or if it ends in a final state symbol within the $\$$ symbols), otherwise it has cost $\gamma=1$ if it results in one of the strings: $\$xp\$$ ($p\notin F$, $x\in \Gamma^*$), $p\$x\$$ ($x\in \Gamma^*$). In the latter cases $\gamma$ has a reduced value because the insertion of $B_1$ (point \ref{it:1}) and a further $3$-substitution are needed to obtain a string that correctly represents the output ID. In any case, a move of $N'$ is simulated by a sequence of transformation $S$, such that $\gamma(S)=3$, and, therefore, a sequence of moves from the initial ID to an accepting one can be simulated with a total cost $3*2^{p(n)}$. At this point, a sequence of $3$-substitutions has to be applied to transform all the symbols within the two $\$$ into $*$. They are $n+1+2^{p(n)}$ at most and each $3$-substitution adds one $*$ at unitary cost. Thus, the whole sequence has cost $\gamma \leq n+1+2^{p(n)}$. Finally, the same cost is required by the sequence of deletions that results in the string $\$\$$.  Thus, $$x \in L\Rightarrow\delta(\$q_0x\$,\$\$)\leq h.$$

    Suppose now $x\notin L$. Any $3$-substitution that does not correspond to a legal move of $N'$, or is part of it, has cost $h+1$, with the exception of those used to transform symbols into $*$ and they can be applied only when the simulated ID is an accepting one. The same holds for deletion of $*$ symbols. Thus, all the sequences of transformations from $\$q_0x\$$ to $\$\$$ have cost $\gamma > n+1+2^{p(n)}$, i.e.: $$x \notin L\Rightarrow\delta(\$q_0x\$,\$\$)> h$$
    
\end{proof}

\begin{theorem}
Any DECIS-$k$-Edit, with $k\geq 3$, is NEXPTIME-complete.
\end{theorem}

\begin{proof}
    The proof is analogue to that of Theorem \ref{th:khamm}. It is easy to demonstrate, by induction, the NEXPTIME-completeness of any DECIS-$k$-Edit problem, with $k\geq 3$, since: a) Algorithm \ref{alg:ntm_3-edit} works for any DECIS-$k$-Edit; b) There exists a polynomial time reduction from DECIS-$k$-Edit to DECIS-$(k+1)$-Edit ($k\geq 2$). Again, the reduction has to inhibit any $(k+1)$-substitution not representing a $k$-substitution.
\end{proof}

\subsection{DECIS-$2$-Edit NEXPTIME-completeness}
\label{sub:2E}
We deal now with the proof of NEXPTIME-completeness of DECIS-$2$-Edit. To prove that DECIS-$2$-Edit $\in$ NEXPTIME, the same algorithm of Section \ref{sec:d3e-nexptime} can be employed (Algorithm \ref{alg:ntm_3-edit}), but, instead of explicitly showing that exists a polynomial time reduction from any problem in NEXPTIME to DECIS-$2$-Edit, we show a polynomial time reduction from a NEXPTIME-complete problem. We, indeed, proved in Section \ref{sec:d3e-nexptime} the NEXPTIME-completeness of DECIS-$3$-Edit, but the same proof is actually valid for a restricted version of the problem, namely DECIS'-$3$-Edit, where, given an instance $<v,w,D,h>$: a) any insertion, deletion or substitution costs either $1$ or $h+1$; b) $3$-substitutions costs are limited to $1$, $3$ or $h+1$.

Therefore, to prove the NEXPTIME-completeness of DECIS-$2$-Edit, it is sufficient to show a reduction from DECIS'-$3$-Edit to it.
\begin{theorem}
There exists a polynomial time reduction from DECIS'-$3$-Edit to DECIS-$2$-Edit.
\end{theorem}

\begin{proof}
The reduction transforms any instance $x= < (v, w, D, h) >$ for DECIS'-$3$-Edit in an instance $x' =< (v, w, D', 5h) >$ for DECIS-$2$-Edit as follows:
\begin{enumerate}
    \item if $\Sigma$ is the alphabet of the input instance, $\Sigma'$ for the output instance is augmented by adding the following new symbols:
    \begin{enumerate}
        \item $S^{i}_{(abc)(def)}$, $\forall a,b,c,d,e,f \in \Sigma, i \in \{1,2,3\}$;
        \item the supporting symbol $*$;
    \end{enumerate}
    \item for each $\epsilon\rightarrow a \in D$ s.t. $\gamma(\epsilon\rightarrow a)=1$, add $\epsilon\rightarrow a$, with $\gamma=5$, in $D'$;
    \item for each $a\rightarrow \epsilon \in D$, s.t. $\gamma (a\rightarrow \epsilon)=1$, add $a\rightarrow \epsilon$, with $\gamma=5$, in $D'$;
    \item for each $a \rightarrow b \in D$ s.t. $\gamma(a\rightarrow b)=1$, add $a\rightarrow b$, with $ \gamma=5$, in $D'$
    \item \label{point:3s} for each $3$-substitution $abc \rightarrow def \in D$ s.t. $\gamma(abc\rightarrow def)=k\leq h$, add the following operations to $D'$:
    \begin{enumerate}
        \item $\epsilon\rightarrow S^{1}_{(abc)(def)}$, with $\gamma=5k-4$;
        \item $aS^{1}_{(abc)(def)}\rightarrow dS^{2}_{(abc)(def)}$, with $\gamma=1$;
        \item $S^{2}_{(abc)(def)}b\rightarrow eS^{3}_{(abc)(def)}$, with $\gamma=1$;
        \item $S^{3}_{(abc)(def)}c\rightarrow f*$, with $\gamma=1$;
        \item $*\rightarrow \epsilon$, with $\gamma=1$;
    \end{enumerate}
    \item any other operation has cost $5h+1$ in $D'.$
\end{enumerate}
The algorithm requires polynomial time. Source and target strings are unchanged, the limit $h$ has to be multiplied by $5$ and the size of the alphabet (and of $D'$) is increased by a polynomial function: $|\Sigma'|=O(|\Sigma|^6)$. 

We can observe that the algorithm is actually a reduction, i.e: $$ x\in \mbox{ DECIS'-}3\mbox{-Edit} \iff x' \in \mbox{ DECIS-}2\mbox{-Edit}$$

Suppose $x=<v,w,D,h>\in \mbox{ DECIS'-}3\mbox{-Edit}$, i.e. $\exists T \ s.t.\ T(v)=w, \gamma(T)\leq h$. Let be $T=t_1t_2\ldots t_n$: it is possible to ``simulate'' each $t_i$ on $x'$ with a sequence of one or more operations $T'_i$ at cost $\gamma(T'_i)=5*\gamma(t_i)$. Insertions, deletions and substitutions require a single operation, while, for $3$-substitutions, the whole sequence of operations described at point \ref{point:3s} is needed, with total cost of $5k$, where $k$ is the original $3$-substitution cost. Therefore, $$x=<v,w,D,h>\in \mbox{ DECIS'-}3\mbox{-Edit}\Rightarrow x'=<v,w,D',5h>\in \mbox{ DECIS-}2\mbox{-Edit}$$

On the other hand, suppose $x=<v,w,D,h>\notin \mbox{ DECIS'-}3\mbox{-Edit}$. It can be observed that each operation on $x'$ either:
\begin{itemize}
\item has cost larger than $5h+1$ and can not be part of an acceptable sequence;
\item corresponds to an operation $t_i$ on $x$, with cost $5*\gamma(t_i)$;
\item is part of a $3$-substitution simulation. Each step of the sequence can be executed only after the previous and the first step introduces a symbol that can not be part of $w'$. The only possibility to remove ``exogenous'' symbols is to apply all the operations in the sequence, at cost  $5*\gamma(t_i)$, where $t_i$ is the simulated 3-substitution.
\end{itemize}

Therefore, $$x=<v,w,D,h>\notin \mbox{ DECIS'-}3\mbox{-Edit}\Rightarrow x'=<v,w,D',5h>\notin \mbox{ DECIS-}2\mbox{-Edit}$$

\end{proof}

\section{Conclusions}\label{sec:conclusions}

In this work we studied the computational complexity of the problems of computing the cost of the  $k$-Hamming  and $k$-Edit distances, for $k\geq 2$, proving that the decision versions that include the description of the distance as part of the instances are, respectively, $\mathbb{P}$-SPACE-complete and NEXPTIME-complete.

We have some preliminary results, not included in this paper, for some special cases where the size of the description of the distance is considered constant. For instance, we found a polynomial time algorithm to compute the $2$-Hamming distance when every operation has the same constant cost.

It is an open problem to find the complexity of solving both problems as the lengths of the two words increase when the distance is fixed, i.e. its size is considered as a constant, or, more generally, when the complexity is further parameterized analogously as done in \cite{Meister15,BarbayP18} for the swap-insert correction distance.

\bibliographystyle{splncs04}
\bibliography{mybibliography}

\end{document}